\documentstyle[aps,prd]{revtex}
\tighten 

\begin{document}
\draft
\twocolumn

\wideabs{
\title{On the size of the smallest scales in cosmic string networks}

\author{Xavier Siemens, Ken D. Olum and Alexander Vilenkin}

\address{Institute of Cosmology\\
Department of Physics and Astronomy\\
Tufts University\\
Medford MA 02155, USA}

\date{\today}

\maketitle

\begin{abstract}
We present a method for the calculation of the gravitational back 
reaction cutoff on the smallest
scales of cosmic string networks taking into account that not all
modes on strings interact with all other modes. This results
in a small scale structure cutoff that is sensitive to the 
initial spectrum of perturbations present on strings. From a simple model, 
we compute the cutoffs in radiation- and matter-dominated universes.
\end{abstract}


}

\section{Introduction}

Topological defects are a generic prediction of grand unified theories.
They can be formed in phase transitions when the topology of the
vacuum manifold of the low energy theory is non-trivial
\cite{kibble76}. For a review see \cite{alexbook}.

The result of a phase transition that produces strings is a
network of long strings that stretch across the horizon and a
collection of closed loops. If the phase transition occurs at
energy \( \eta_{s} \), then the mass per unit length in the
strings is \( \mu \sim \eta ^{2}_{s} \). Immediately after formation, 
strings in the early universe undergo an
epoch of heavy damping until a time $t_*$, which depends on the
string formation scale  \( \eta_{s} \) as well as the specifics of the
particle physics model that
produces them. As a result of this damping all sub-horizon
features, such as loops and perturbations on long strings, are
wiped out and the long strings and loops are Brownian with a
persistence length $d\sim t_*$. 

After that time,
numerical simulations
show that in an expanding universe there is an attractor solution 
in which the network
evolves into a ``scaling regime'' (see \cite{alexbook}
and references therein), where the energy density of the string
network is a small constant fraction of the radiation or matter density
and the statistical properties of the system, such as the
correlation lengths of long strings and average sizes of loops,
scale with the cosmic time \( t \).  This solution is possible because of
intercommutations that produce cosmic string 
loops which 
in turn decay by radiating gravitationally. Simulations also 
have found that most 
loops and the perturbations on long strings have the smallest possible size,
the simulation resolution, which does not scale. 
The prevailing opinion on this
issue is that the size of small-scale structure in fact also scales with
the cosmic time \( t \) and its value is given by the
gravitational back-reaction scale \( \Gamma G\mu t \), where \(
\Gamma  \) is a number of order 100 and \( G \) is Newton's
constant. This possibility was first pointed out in \cite{BB}.

Cosmic strings are good candidates for a variety of interesting 
cosmological phenomena such as gamma ray bursts \cite{4}, 
gravitational waves \cite{5,AllenOtt} and ultra high energy cosmic 
rays \cite{6,Bere}. Some
of the predictions of these models, however, depend sensitively on the so
far unresolved question of the size of the small-scale structure.

\section{Back-reaction model}

Hindmarsh \cite{hindmarsh} (see also \cite{battye}) 
showed that the power per unit length
radiated into gravitational waves from two colliding arbitrary
small perturbations of long repeat length $L$ on an infinite
string is
\begin{equation}
\label{hindapprox}
{dP \over dl} \sim \pi G \mu^2  \sum_{n,m}
(\kappa_n+\kappa_m)\epsilon^2 _n \epsilon^2 _m,
\end{equation}
where $\kappa_n=2\pi n/L$ and
$\kappa_m=2\pi m/L$ are the wavenumbers of the right- and
left-moving Fourier modes that make up the arbitrary perturbations
and $\epsilon_n$ and $\epsilon_m$ their (small) amplitude to wavelength
ratios.

We can use this expression to construct a simple back-reaction
model. If we split the sum of the wavenumbers in
Eq. (\ref{hindapprox})
\begin{equation}
\label{pownm} {dP \over dl} = \sum_{n}{dP_n \over dl}+
\sum_{m}{dP_m \over dl},
\end{equation}
we can identify individual modes with the power they radiate
\begin{equation}
\label{pown}
{dP_n \over dl} \sim \pi G \mu^2  \epsilon^2 _n
\kappa_n \sum_{m} \epsilon^2 _m,
\end{equation}
with a similar expression for the $m$ modes travelling in the
opposite direction. Each of the modes contributes to the effective
mass per unit length of the string by an amount
\begin{equation}
\label{dmun}
\delta \mu_n \approx \mu \epsilon^2 _n
\end{equation}
and the power radiated into
gravitational waves by each of the modes decreases this
contribution \cite{hindmarsh}
\begin{equation}
\label{dpowdmu}
{dP_n \over dl} = - {d \over dt} (\delta \mu_n).
\end{equation}
By putting Eq. (\ref{dpowdmu}) together with Eqs. (\ref{dmun}) and
(\ref{pown}) we arrive at an expression for the time evolution of
the amplitude to wavelength ratio of each of the modes
\begin{equation}
{\dot \epsilon}_n \sim -\pi G \mu \kappa_n \epsilon_n \sum_m
\epsilon^2 _m
\label{epseq}
\end{equation}
with an analogous expression for the left-moving $m$ modes. Each mode
therefore loses amplitude at a rate proportional to its own
frequency, to its amplitude to wavelength ratio and to a sum which
represents the interaction of the mode with every other mode moving in
the opposite direction.

There remains, however, some puzzling behavior. In the case of
just two colliding modes we can write Eq. (\ref{hindapprox}) as
\begin{equation}
\label{hindapprox2modes}
{dP \over dl} \sim \pi G \mu^2
(\kappa_a+\kappa_b)\epsilon^2 _a \epsilon^2 _b
\end{equation}
where the $a$ and $b$ subscripts differentiate between the right-
and left-moving modes. If we conformally stretch (say) the $a$
mode such that $\epsilon_a$ remains constant while $\kappa_a
\rightarrow 0$, Eq. (\ref{hindapprox2modes}) approaches the constant
\begin{equation}
\label{hindapprox2modesconst}
{dP \over dl} \sim \pi G \mu^2
\kappa_b\epsilon^2 _a \epsilon^2 _b.
\end{equation}
This result is in contradiction to the fact that the power
vanishes in the case of perturbations travelling in only one
direction on a string \cite{Vachaspati}: By conformally stretching
one of the modes we are making the string on which the other mode
is travelling straighter and we expect the power radiated to
approach zero in that limit.

This problem was investigated in \cite{us}. Here we will summarise 
the results. It turns out that in order to be in the Hindmarsh 
regime (the regime where  Eq. (\ref{hindapprox2modes}) is valid) it is not
merely sufficient for the amplitudes of each of the colliding modes
to be smaller 
than their corresponding wavelengths: 
It is also necessary for both amplitudes
to be small compared to the geometric 
mean of the wavelengths, namely it is necessary to have
\begin{equation}
\label{conda1}
\epsilon_a ^2 {\kappa_b \over \kappa_a}={A_a ^2 \over \lambda_a
\lambda_b} \ll 1
\end{equation}
and
\begin{equation}
\label{conda2}
\epsilon_b ^2 {\kappa_a \over \kappa_b}={A_b ^2 \over \lambda_a
\lambda_b} \ll 1,
\end{equation}
where $A_a$ and $A_b$ are the amplitudes 
and $\lambda_a$ and
$\lambda_b$ are the wavelengths of the two colliding modes.
If instead we are in the regime where one of these two Lorentz 
invariant quantities, say,
\begin{equation}
\label{cond}
\epsilon_a ^2 {\kappa_b \over \kappa_a}={A_a ^2 \over \lambda_a
\lambda_b} \gg 1
\end{equation}
the power radiated is 
exponentially suppressed. This is precisely the regime
where the power given by Eq. (\ref{hindapprox2modes}) tends to the
constant Eq. (\ref{hindapprox2modesconst}). It was also found 
that when the quantity
$\epsilon_a ^2 \kappa_b / 8\kappa_a$ increases past $1$ 
the power drops discontinuously. 

We can approximate this behaviour by introducing a cutoff 
in Eq. (\ref{epseq}) such that the sum excludes interactions 
between modes that satisfy
\begin{equation}
\label{cond2b}
\epsilon_a ^2 {\kappa_b \over \kappa_a} > 8,
\end{equation}
as follows
\begin{equation}
\label{backreact}
{\dot \epsilon}_n \sim -\pi G \mu \kappa_n \epsilon_n \sum_m
\epsilon^2 _m \theta (8-\epsilon^2 _n {\kappa_m \over
\kappa_n})\theta (8-\epsilon^2 _m {\kappa_n \over \kappa_m})
\end{equation}
with a similar equation for the $m$-modes traveling in the opposite direction.
This approximation is justified because of the discontinuous drop
in the power that takes place when $\epsilon_a ^2 \kappa_b / \kappa_a = 8$
and the exponential 
suppression that sets in when Eq. (\ref{cond}) is satisfied. In fact
we cannot easily associate the power radiated with a given mode outside
of the Hindmarsh regime and it is fortunate that in this case the power 
radiated is exponentially supressed and therefore negligible. 
It should be noted that the calculation 
in \cite{us} was performed for just two modes travelling in 
opposite directions. Here we have 
assumed (reasonably, we believe) that when a mode interacts 
with a collection of modes all of which satisfy Eq. (\ref{cond}) the power
is also negligible.

Eq. (\ref{backreact}) can be used to find the evolution of a set
of amplitude to wavelength ratios from some initial conditions.  There
are two distinct cases of interest.  On a long string, one can
consider Eq. (\ref{backreact}) to give the damping of the average
perturbation with a certain wavelength.  By symmetry the left-moving
and right-moving perturbations will have the same average amplitudes.
However, in the case of a loop, a statistical fluctuation may produce
an excess of power in a particular direction over a broad range of
wavelengths.  In that case, since a perturbation passes the same
oppositely-directed perturbations over and over again, it is essential
to treat the right- and left-movers separately. Here we will consider
the case of long strings only.
  
\section{The size of the smallest scales}

When the time-scale over which modes on a long string are significantly 
stretched by the expansion of the universe, which 
is about $t$, the cosmic time, is large compared to the gravitational
damping time, Eq. (\ref{backreact}) can be used to
estimate the size of the smallest scales on long strings: 
The $n$th mode will be
significantly damped in amplitude provided $\tau_g < t$ with $\tau_g$
given by
\begin{equation}
\label{taug} {{\dot \epsilon}_n \over \epsilon_n} \sim -{1
\over \tau_g}.
\end{equation}

We expect that a combination of stretching by the expansion 
of the universe and self-intersections on strings will result
in an average spectrum of pertubations where the small scales 
are suppressed relative to the large scales. This seems reasonable 
even when kinks (produced at intercommutation events) are present 
since the amplitude to wavelength 
ratio of Fourier components of kinks as a function of mode number
decays like $1/n$.
The presence of a flat or decreasing spectrum  
means the sum in Eq. (\ref{backreact}) is
going to be bounded from below and above: A given mode does not
see every other mode on the string but rather a range of
modes that lie around it. If we assume a power law spectrum
\begin{equation}
\label{spec} 
\epsilon_n \sim n^{-\beta},
\end{equation}
one can use Eq. (\ref{cond2b}) to show that the lower bound on $m$ for a given
mode $n$ is
\begin{equation}
\label{mMin} 
m_{\text{Min}} \sim \left( {n \over 8} \right)^{1/(1+2\beta)},
\end{equation}
which corresponds to the largest wavelength mode that the $n$th mode
interacts with while still in the Hindmarsh regime. The smallest
wavelength mode that it interacts with can also be found from Eq. (\ref{cond2b})
and is given by
\begin{equation}
\label{mMax} m_{\text{Max}} \sim 8n^{1+2\beta}.
\end{equation}
We can therefore write Eq. (\ref{backreact}) as
\begin{equation}
\label{backreactMinMax} {{\dot \epsilon}_n \over \epsilon_n} \sim
-\pi G \mu \kappa_n \sum_{m=m_{\text{Min}}} ^{m_{\text{Max}}} \epsilon^2 _m
\end{equation}
with $m_{\text{Min}}$ and $m_{\text{Max}}$ given by Eqs. (\ref{mMin}) and
(\ref{mMax}) above.

Previously, it was assumed that all modes interact with all other
modes and therefore for spectra where the small scales are sufficiently 
suppressed, namely those with $\beta > 1/2$,
\begin{equation}
\label{sumeq1} \sum_{m} \epsilon^2 _m \sim 1.
\end{equation}
This means that we would expect the survival of modes whose
wavelengths
\begin{equation}
\label{SSCprev} \lambda > 2 \pi^2 G \mu t,
\end{equation}
independently of the specific spectrum of perturbations present 
on the string.

In fact we need to evaluate the sum between the two limits. 
We can approximate the sum in Eq. (\ref{backreactMinMax}) by an
integral
\begin{equation}
\label{sumeqsomething} 
\sum_{m=m_{\text{Min}}} ^{m_{\text{Max}}} \epsilon^2 _m
\sim \int_{m_{\text{Min}}} ^{m_{\text{Max}}}  dm m^{-2\beta}
\end{equation}
and take the dominant contribution which when $\beta > 1/2$ is given by
\begin{equation}
\label{summm} 
\int_{m_{\text{Min}}} ^{m_{\text{Max}}}  dm {m^{-2\beta}} 
\sim -{1 \over 1-2\beta}{m_{\text{Min}}} ^{1-2\beta}.
\end{equation}
The dominant contribution therefore comes from the
largest wavelength mode that the $n$th mode can interact with.

Using Eq. (\ref{summm}) we can write Eq. (\ref{backreactMinMax}) as
\begin{equation}
\label{backreactSumDone} {{\dot \epsilon}_n \over \epsilon_n} \sim
{1 \over 1-2\beta} \pi G \mu \kappa_n {m_{\text{Min}}} ^{1-2\beta}.
\end{equation}
Using Eqs. (\ref{taug}) and (\ref{mMin}) we can see that
\begin{equation}
\label{trans1} {1 \over \tau_g} \sim { \Gamma G \mu \over \lambda}
n^{(1-2\beta)/(1+2\beta)},
\end{equation}
where 
\begin{equation}
\label{Gamma}
\Gamma={8^{(2\beta-1)/(2\beta+1)} \over 2\beta-1 } 2\pi^2 .
\end{equation}
Since we want
the gravitational lifetime of the mode to be larger than the
cosmic time and using $n \sim t/\lambda$,
\begin{equation}
\label{trans2} { \Gamma G \mu \over \lambda} \left({t \over
\lambda}\right)^{(1-2\beta)/(1+2\beta)} < {1 \over t},
\end{equation}
and therefore only modes with wavelengths
\begin{equation}
\label{cutoff} \lambda > \left(\Gamma G \mu \right)^{(1+2\beta)/2} t,
\end{equation}
have a significant amplitude at time $t$. 

\section{Simple Examples}

When the amplitude is small
compared to the wavelength, one can find the effect
of cosmological expansion on the spectrum of perturbations on a string
\cite{alexstretch}. 
If the wavelength of a mode on a
long string is larger than the horizon, then it is stretched
conformally; its amplitude and wavelength grow with the scale
factor. If, on the other hand, the wavelength of the mode is
smaller then the horizon, only its wavelength is stretched. The
net result of this cosmological processing is that the amplitude
to wavelength ratio of modes larger than the horizon as a function
of the wavenumber remains constant and for modes inside the
horizon it becomes a power law.

It is not hard to see why this is so.  If a mode enters the
horizon at a time $t_0$, then its wavelength and amplitude are both
$\sim t_0$. While inside the horizon the amplitude remains fixed
and the wavelength is stretched so that at sometime $t$ later
\begin{equation}
\label{Lstretch} \lambda_0 (t)\sim t_0 \left( {t \over t_0}
\right)^\alpha = t_0 ^{1-\alpha} t^\alpha,
\end{equation}
where $\alpha=1/2$ or $2/3$ depending on whether we are in the
radiation or matter era. We can therefore write the amplitude to
wavelength ratio of that mode as
\begin{equation}
\label{epsilon} \epsilon_0 (t) \sim {t_0 \over t_0 ^{1-\alpha}
t^\alpha} = {t_0 ^\alpha \over t^\alpha}.
\end{equation}
If we imagine that our mode of wavelength $\lambda_0 (t)$ is some
fraction of the largest mode, $\lambda_0 (t) = \lambda /n $ with
$\lambda \sim t$, it is easy to see that
\begin{equation}
\label{L0} t_0 = {t  n^{-{1 / (1-\alpha)}}}
\end{equation}
which when substituted into Eq. (\ref{epsilon}) yields the spectrum
\begin{equation}
\label{epsilonN} 
\epsilon_n \sim {n^{-{\alpha / (1-\alpha)}}}=
\left\{
\begin{array}{cc}
n^{-1} & \text{Radiation Era} \\
n^{-2} & \text{Matter Era.}
\end{array}
\right.
\end{equation}
Feeding the spectrum Eq. (\ref{epsilonN}) into Eq. (\ref{cutoff}) 
yields
\begin{equation}
\label{cutoff2} \lambda > \left(\Gamma G \mu \right)^{(1+\alpha)/
 2(1-\alpha)} t=
\left\{
\begin{array}{cc}
\left( \Gamma G \mu \right)^{3/2} t& \text{Radiation Era}\\
\left( \Gamma G \mu \right)^{5/2} t& \text{Matter Era}
\end{array}
\right.
\end{equation}
as the small scale structure cutoffs in the radiation and matter eras. 
However, this analysis neglects 
intercommutations, which produce kinks,
and the possibility that small perturbations propagating on curved 
horizon-sized strings may not be efficiently stretched.

\section{Conclusions}

We have presented a method for the calculation of gravitational
back-reaction on cosmic strings taking into account
that not all Fourier modes that make up the perturbations on strings
interact with all other modes with the same 
efficiency\cite{us}. 

In particular, modes
of a given wavelength only interact significantly with a narrow range 
of other modes whose wavelengths are comparable to it. This range depends 
on the spectrum
of perturbations on the string at late times (when gravitational effects 
become important) which in turn has the effect of making
the small scale structure cutoff sensitive to this spectrum.
 
Using this method and assuming a power law spectrum 
we have arrived at an expression for the small scale structure cutoff.
We have further applied it to the spectrum resulting from the stretching of 
small amplitude waves in a radiation- or matter-dominated FRW universe.

It is unclear to what extent
the results obtained for these two simple cases apply to a realistic 
network of cosmic strings because
we have ignored
the effect of intercommutations, which leads to kinks, as well as the 
possibility that small perturbations propagating on curved horizon-sized
strings may not be significantly stretched by the expansion of the universe. 
Both of these effects have the potential to
change the spectrum of perturbations
and therefore the value of the small scale structure cutoff. 

However, it is clear that because of the restricted range of interaction
affecting every mode, the small scale structure cutoff will be smaller 
than the one given by the usual calculation where an 
unrestricted range of interaction is assumed.

\acknowledgments
We would like to thank Jose Juan Blanco-Pillado for useful discussions.
The work of KDO and AV was partially funded by the NSF.

\end{document}